\documentclass[pra,twocolumn,superscriptaddress,showpacs,nofootinbibfloatfix,amsmath,amsfonts,amssymb]{revtex4-1}%
\usepackage{amsmath,amsfonts,amssymb,color}
\usepackage{amsthm}
\usepackage{leftidx}
\usepackage{graphicx}
\usepackage{xcolor}
\usepackage{dcolumn}
\usepackage{bm}
\usepackage{epstopdf}
\usepackage{epsfig}
\usepackage{environ}
\usepackage{pdfcomment}

\usepackage{float}
\usepackage[T1]{fontenc}
\usepackage[latin9]{inputenc}
\usepackage{setspace}
\usepackage{esint}

\begin{document}

\title{Floquet dynamical quantum phase transitions in periodically quenched systems}

\author{Longwen Zhou}
\email{zhoulw13@u.nus.edu}
\affiliation{%
	Department of Physics, College of Information Science and Engineering, Ocean University of China, Qingdao, China 266100
}

\author{{Qianqian Du}}
\affiliation{%
	Department of Physics, College of Information Science and Engineering, Ocean University of China, Qingdao, China 266100
}

\date{\today}

\begin{abstract}
Dynamical quantum phase transitions (DQPTs) are characterized by nonanalytic
behaviors of physical observables as functions of time. When a system
is subject to time-periodic modulations, the nonanalytic signatures of its
observables could recur periodically in time, leading to the phenomena
of Floquet DQPTs. In this work, we systematically explore Floquet
DQPTs in a class of periodically quenched one-dimensional system with chiral
symmetry. By tuning the strength of quench, we find multiple Floquet
DQPTs within a single driving period, with more DQPTs being observed when the system is initialized in Floquet states with larger topological invariants. Each Floquet DQPT is further accompanied
by the quantized jump of a dynamical topological order parameter,
whose values remain quantized in time if the underlying Floquet system
is prepared in a gapped topological phase.
The theory is demonstrated in a piecewise quenched lattice
model, which possesses rich Floquet topological phases and is readily
realizable in quantum simulators like the nitrogen-vacancy center
in diamonds. Our discoveries thus open a new perspective for the Floquet
engineering of DQPTs and the dynamical detection of topological phase
transitions in Floquet systems.
\end{abstract}

\pacs{}
\keywords{}
\maketitle

\section{Introduction}\label{sec:Int}

Dynamical quantum phase transitions are featured by the nonanalyticity
of physical observables in time domain~\cite{DQPT1,DQPT2}. They are usually triggered
by a sudden~\cite{DQPT3,DQPT4,DQPT5,DQPT6,DQPT7,DQPT8,DQPT9,DQPT10,DQPT11} or a slow~\cite{DQPTSL1,DQPTSL2} quench across an equilibrium quantum phase transition
point, and will exhibit topological signatures when the pre- and
post-quench systems are set in different topological phases~\cite{DTOP1,DTOP2,DTOP3,DTOP4,DTOP5,DTOP6}. In the
past few years, various aspects of DQPTs have been explored in theory~(see \cite{DQPTRev1,DQPTRev2,DQPTRev3,DQPTRev4,DQPTRev5} for reviews), and also demonstrated experimentally in
a broad range of quantum simulator platforms~\cite{DQPTExp1,DQPTExp2,DQPTExp3,DQPTExp4,DQPTExp5,DQPTExp6,DQPTExp7,DQPTExp8,DQPTExp9,DQPTExp10,DQPTExp11}, yielding an intriguing
perspective in the classification and characterization of nonequilibrium
phases of matter.

Recently, the study of DQPTs has been extended to time-periodic driven
systems~\cite{DQPTExp11,FDQPT2,FDQPT3}, which possess rich nonequilibrium phases like the Floquet
topological matter~\cite{FTP1,FTP2,FTP3,FTP4,FTP5,FTP6} and time crystals~\cite{DTC1,DTC2,DTC3,DTC4,DTC5}. Following periodic driving fields,
a different class of DQPT unique to Floquet systems has been discovered
theoretically and also observed experimentally in a harmonically driven
spin chain, which is thus named Floquet DQPTs~\cite{DQPTExp11}. The nonanalytic
cusps in the return rate of Floquet DQPTs are found to recur
periodically in time with a non-decaying global profile, which is
distinguished from the conventional DQPTs that are usually observable
only in transient time scales~\cite{DQPT1}. Moreover, when the periodically
driven system is initialized in a Floquet topological phase, the Floquet
DQPTs would also show topologically nontrivial signatures, characterized
by the quantized jump of a dynamical topological order parameter (DTOP)
at each critical time of the transition~\cite{DQPTExp11}. In later studies,
different aspects of Floquet DQPTs have also been explored, such as
the effect of mixed state and the property of entanglement spectrum~\cite{FDQPT4,FDQPT5}.
However, all the studies on Floquet DQPTs till now focus on
a specific type of driving protocol, in which the system can be described
by a static effective Hamiltonian in a rotating frame~\cite{DQPTExp11,FDQPT4,FDQPT5}. Therefore,
richer signatures of Floquet DQPTs under more general driving schemes
have not been revealed. For example, is it possible to induce several
Floquet DQPTs within a single driving period, making the nonanalytic
cusps in the rate function of return probability to form a ``\emph{sublattice
	structure in time}''? Besides, are there any differences in DQPTs
between systems initialized in Floquet phases with small and large
topological invariants? Moreover, compared to the cases following
a sudden or a slow quantum quench, much less about DQPTs is known
when the quenches are applied over finite time durations~\cite{DQPTFT1,DQPTFT2}. The resolution
of these issues would further uncover the intrinsic properties of
Floquet DQPTs, and establish their connections with Floquet topological
matter and other nonequilibrium dynamical phenomena.

Motivated by these considerations, we explore Floquet DQPTs in a general
class of periodically quenched lattice model, which possesses the
chiral (sublattice) symmetry at each instant of time. In Sec.~\ref{sec:The},
we introduce our theoretical framework, which shows that multiple
Floquet DQPTs can indeed emerge within a single driving period under
large quench amplitudes. Furthermore, a DTOP for Floquet DQPTs can
be constructed from the noncyclic geometric phase of the evolving
Floquet states~\cite{DQPTExp11}. The DTOP shows a quantized jump whenever a Floquet
DQPT happens, and remains to be an integer otherwise if the system
is initially prepared in a gapped Floquet topological phase. Specially,
the DTOP could take non-quantized values in certain time domains if
the initial state of the system belongs to a gapless Floquet phase.
One can therefore employ the DTOP to identify the boundaries between
different Floquet topological phases. In Sec.~\ref{sec:Res}, we demonstrate
our theory in a prototypical and experimentally realizable periodically
quenched lattice model, unveiling the rich signatures of Floquet DQPTs
by detailed analytical and numerical calculations. In Sec.~\ref{sec:Sum},
we summarize our results and discuss potential future directions.

\section{Theory}\label{sec:The}
In this section, we introduce our theory for the description of Floquet
DQPTs in a typical class of periodically quenched system. Throughout
our discussion, we will set the Planck constant $\hbar=1$ and the
driving period $T=2$ in dimensionless units.

\subsection{Setup and Floquet dynamics}\label{subsec:Setup}
In momentum presentation, the time-dependent Hamiltonian $H(k,t)$
of our periodically quenched lattice takes the form
\begin{equation}
H(k,t)=\begin{cases}
h_{x}(k)\sigma_{x} & t\in[2\ell,2\ell+1),\\
h_{y}(k)\sigma_{y} & t\in[2\ell+1,2\ell+2).
\end{cases}\label{eq:Hkt}
\end{equation}
Here $k\in[-\pi,\pi)$ is the quasimomentum defined in the first Brillouin
zone. $\ell\in\mathbb{Z}$ counts the number of driving periods. $\sigma_{x,y,z}$
are Pauli matrices in their usual representations, and we will denote
the $2\times2$ identity matrix as $\sigma_{0}$. It is clear that
$H(k,t)$ is periodic in time with the period $T=2$, i.e., $H(k,t)=(k,t+2)$.
We will also use the symbol $s$ to denote the time within a driving
period, i.e., the micromotion time, so that any physical time $t$
can be decomposed as 
\begin{equation}
t=s+2\ell,\qquad\ell\in\mathbb{Z},\qquad s\in[0,2).\label{eq:tls}
\end{equation}
Meanwhile, according to Eq.~(\ref{eq:Hkt}), $H(k,t)$ has the chiral
symmetry $\Gamma=\sigma_{z}$ at any instant of time, i.e., $\Gamma H(k,t)\Gamma=-H(k,t)$
and $\Gamma^{2}=\sigma_{0}$. This allows one to characterize the
topological phases of the system by integer winding numbers~\cite{AsbothSTF1,ZhouSQKR,ZhouDSCL}.
Note in passing that Eq.~(\ref{eq:Hkt}) represents one of the typical quench protocols for the realization of one-dimensional~(1D) Floquet systems with chiral (sublattice) symmetry and nontrivial topological properties. The Pauli matrices in Eq.~(\ref{eq:Hkt}) can be replaced by any two of the three Pauli spins $\sigma_{x,y,z}$, and $h_{x,y}$ can be arbitrary functions of the quasimomentum $k$. Various other periodic kicking or quenching protocols for 1D chiral symmetric lattice models can be reduced to the form of Eq.~(\ref{eq:Hkt}), such as those reported in Refs.~\cite{AsbothSTF1,ZhouSQKR}. Therefore, the theoretical framework developed here is expected to be appliable to Floquet DQPTs in a broad range of periodically quenched and continuously driven~\cite{DQPTExp11} chiral symmetric two-band systems.

We are interested in such a dynamical process, in which the system is
initialized in a uniformly filled Floquet band and then undergoes
nonequilibrium evolution governed by the piecewise quenched $H(k,t)$.
The initial state of the system is obtained by solving the Floquet
eigenvalue equation $U(k)|\psi(k)\rangle=e^{-iE(k)}|\psi(k)\rangle$,
where $|\psi(k)\rangle$ is the Floquet eigenstate with eigenphase
$E(k)$. The Floquet operator $U(k)$, referring to the evolution
operator of the system over a complete driving period $T$ (e.g.,
from $t=0^{-}$ to $t=0^{-}+2$), takes the form
\begin{equation}
U(k)=e^{-ih_{y}(k)\sigma_{y}}e^{-ih_{x}(k)\sigma_{x}},\label{eq:Uk}
\end{equation}
according to Eq.~(\ref{eq:Hkt}). Taking the Taylor expansion of the
two terms on the right hand side of Eq.~(\ref{eq:Uk}), we can express
$U(k)$ equivalently as
\begin{alignat}{1}
U(k)& = \cos[h_{x}(k)]\cos[h_{y}(k)]\sigma_{0}\nonumber \\
& -i \sin[h_{x}(k)]\cos[h_{y}(k)]\sigma_{x}\nonumber \\
& -i \sin[h_{y}(k)]\cos[h_{x}(k)]\sigma_{y}\nonumber \\
& +i \sin[h_{x}(k)]\sin[h_{y}(k)]\sigma_{z},\label{eq:Ukep}
\end{alignat}
which can be further recast into a single exponential form as
\begin{equation}
U(k)=e^{-iE(k)\vec{n}(k)\cdot\vec{\sigma}}=e^{-i\vec{d}(k)\cdot\vec{\sigma}}.\label{eq:Ukn}
\end{equation}
Here the eigenphase dispersion reads
\begin{equation}
E(k)=\arccos\{\cos[h_{x}(k)]\cos[h_{y}(k)]\},\label{eq:Ek}
\end{equation}
$\vec{\sigma}=(\sigma_{x},\sigma_{y},\sigma_{z})$, and the components
of unit vector $\vec{n}=[n_{x}(k),n_{y}(k),n_{z}(k)]$ are explicitly
given by
\begin{alignat}{1}
n_{x}(k)& = \sin[h_{x}(k)]\cos[h_{y}(k)]/\sin[E(k)],\nonumber \\
n_{y}(k)& = \sin[h_{y}(k)]\cos[h_{x}(k)]/\sin[E(k)],\label{eq:nk}\\
n_{z}(k)& = -\sin[h_{x}(k)]\sin[h_{y}(k)]/\sin[E(k)].\nonumber 
\end{alignat}
We could then obtain the Floquet eigenstates by solving the eigenvalue
equation of the effective Floquet Hamiltonian, i.e.,
\begin{equation}
H_{{\rm eff}}(k)|\psi_{\pm}(k)\rangle=E_{\pm}(k)|\psi_{\pm}(k)\rangle,\label{eq:HeffEq}
\end{equation}
where the effective Hamiltonian $H_{{\rm eff}}(k)$ and eigenphase
bands $E_{\pm}(k)$ are given by
\begin{equation}
H_{{\rm eff}}(k)=\vec{d}(k)\cdot\vec{\sigma},\qquad E_{\pm}(k)=\pm E(k),\label{eq:HeffE}
\end{equation}
and the explicit expressions of Floquet eigenstates $|\psi_{\pm}(k)\rangle$
are found to be 
\begin{equation}
|\psi_{v}(k)\rangle=\frac{1}{\sqrt{2E_{v}(k)[E_{v}(k)-d_{z}(k)]}}\begin{bmatrix}d_{x}(k)-id_{y}(k)\\
E_{v}(k)-d_{z}(k)
\end{bmatrix},\label{eq:Psivk}
\end{equation}
where $v=\pm$ will be used to denote the signs of the eigenphase
dispersions $E_{\pm}(k)$. With these considerations, the time evolution
operator of the system from time $t=0$ to a later time $t=s+2\ell$ {[}see
Eq.~(\ref{eq:tls}){]} can be written as $U(k,t)=U(k,s)U^{\ell}(k)$,
where
\begin{equation}
U(k,s)=\begin{cases}
e^{-ish_{x}(k)\sigma_{x}}, & s\in[0,1)\\
e^{i(2-s)h_{y}(k)\sigma_{y}}U(k). & s\in[1,2)
\end{cases}\label{eq:Uks}
\end{equation}
Here $U(k)$ is the Floquet operator of the system over a complete driving period, as given by Eq.~(\ref{eq:Uk})~\cite{Note1}.
The evolution of a Floquet eigenstate $|\psi_{v}(k)\rangle$ ($v=\pm$)
from $t=0$ to $t=s+2\ell$ is then described by $|\psi_{v}(k,t=s+2\ell)\rangle=U(k,s)U^{\ell}(k)|\psi_{v}(k)\rangle$,
which can be expressed more explicitly as
\begin{equation}
|\psi_{v}(k,t)\rangle=e^{-i\ell E_{v}(k)}e^{-ish_{x}(k)\sigma_{x}}|\psi_{v}(k)\rangle\label{eq:Psivkt1}
\end{equation}
for $s\in[0,1)$ and 
\begin{equation}
|\psi_{v}(k,t)\rangle=e^{-i(\ell+1) E_{v}(k)}e^{i(2-s)h_{y}(k)\sigma_{y}}|\psi_{v}(k)\rangle\label{eq:Psivkt2}
\end{equation}
for $s\in[1,2)$, respectively.
Since $\{|\psi_{v}(k)\rangle|v=\pm\}$ in Eq.~(\ref{eq:Psivk}) form a complete and orthonormal
basis, the dynamics of our periodically quenched system is completely
solved by Eqs.~(\ref{eq:Psivkt1})-(\ref{eq:Psivkt2}) once the explicit expressions of $h_{x}(k)$
and $h_{y}(k)$ are specified.

\subsection{Return amplitude and conditions of Floquet DQPTs}\label{subsec:RA}
With these preparations, we can now investigate the Floquet DQPTs
in our class of piecewise quenched system. The central object in
the study of DQPTs is the return amplitude (or return probability),
which may be viewed as a dynamical version of the partition function~\cite{DQPTRev1}.
For an evolution from time $t=0$ to $t=s+2\ell$ {[}see Eq.~(\ref{eq:tls}){]},
the return amplitude of a single Floquet-Bloch state $|\psi_{v}(k)\rangle$
($v=\pm$) is given by 
\begin{equation}
G_{v}(k,t)=\langle\psi_{v}(k)|U(k,t)|\psi_{v}(k)\rangle=e^{-i\ell E_{v}(k)}G_{v}(k,s),\label{eq:Gvkt}
\end{equation}
where
\begin{equation}
G_{v}(k,s)=\langle\psi_{v}(k)|U(k,s)|\psi_{v}(k)\rangle\label{eq:Gvks}
\end{equation}
describes the return amplitude of initial state $|\psi_{v}(k)\rangle$
at a given $k\in[-\pi,\pi)$ in micromotion time $s\in[0,2)$. Following the theory
of Floquet DQPTs~\cite{DQPTExp11}, if a state $|\psi_{v}(k)\rangle$ with $k=k_{c}$
happens to evolve to its orthogonal state at a critical time $s=s_{c}$,
we will have $G_{v}(k_{c},s_{c})=0$, and the rate function of return
probability (to be defined later) will become nonanalytic in the thermodynamic
limit, signifying Floquet DQPTs at time $t_{c}=s_{c}+2\ell$ for
all $\ell\in\mathbb{Z}$. 

The conditions of Floquet DQPTs are determined by the expressions
of critical momentum $k_{c}$ and critical time $t_{c}$. For our
periodically quenched system, the forms of $k_{c}$ and $t_{c}$ can
be found analytically. Using Eqs.~(\ref{eq:Psivkt1})-(\ref{eq:Gvks})
and the fact that $\langle\psi_{v}(k)|\sigma_{j}|\psi_{v}(k)\rangle=vn_{j}(k)$
for $j=x,y,z$ [see Eq.~(\ref{eq:nk}) for $n_{x,y,z}(k)$], we find that
\begin{equation}
G_{v}(k,s)=\begin{cases}
\cos[sh_{x}(k)]-i\sin[sh_{x}(k)]vn_{x}(k),\\
e^{-iE_{v}(k)}\{\cos[s'h_{y}(k)]+i\sin[s'h_{y}(k)]vn_{y}(k)\},
\end{cases}\label{eq:Gvks2}
\end{equation}
where $s\in[0,1)$, $s'\equiv2-s\in[0,1)$ and $[n_{x}(k),n_{y}(k)]$ are the components
of unit vector $\vec{n}(k)$ in Eq.~(\ref{eq:nk}). The expressions
of $k_{c}$ and $s_{c}$ are then determined by the solutions of $G_{v}(k,s)=0$,
yielding the following two sets of critical conditions
\begin{equation}
\begin{cases}
sh_{x}(k)=\frac{2p-1}{2}\pi\quad\&\quad\sin[h_{x}(k)]\cos[h_{y}(k)]=0,\\
s'h_{y}(k)=\frac{2q-1}{2}\pi\quad\&\quad\sin[h_{y}(k)]\cos[h_{x}(k)]=0,
\end{cases}\label{eq:CriCd}
\end{equation}
for $s\in[0,1)$ and $s'\equiv2-s\in[0,1)$. Here $p,q\in\mathbb{Z}$ and Eq.~(\ref{eq:nk}) has been used to reach
the second equality in each set of condition. Note that these conditions
do not depend on the values of $v$~($=\pm$), implying that we would have
the same Floquet DQPT conditions for the initial state being prepared in
either Floquet band. By straightforward calculations, we see that
in each half of the driving period, the condition in Eq.~(\ref{eq:CriCd})
generates two possible sets of solutions for the critical momenta
and time. For $s\in[0,1)$, we have $G_{v}(k_{c},s_{c})=0$
if
\begin{equation}
\begin{cases}
h_{x}(k_{c})=m\pi, & m\in\mathbb{Z}_{\neq0}\\
s_{c}=\frac{2p-1}{2m}\in(0,1), & p\in\mathbb{Z}
\end{cases}\label{eq:kcsc11}
\end{equation}
or 
\begin{equation}
\begin{cases}
h_{y}(k_{c})=\frac{2m-1}{2}\pi, & m\in\mathbb{Z}\\
s_{c}=\frac{2p-1}{2h_{x}(k_{c})}\in(0,1). & p\in\mathbb{Z}
\end{cases}\label{eq:kcsc12}
\end{equation}
For $s\in[1,2)$, we have $G_{v}(k_{c},s_{c})=0$ if
\begin{equation}
\begin{cases}
h_{y}(k_{c})=n\pi, & n\in\mathbb{Z}_{\neq0}\\
s'_{c}=\frac{2q-1}{2n}\in(0,1), & q\in\mathbb{Z}
\end{cases}\label{eq:kcsc21}
\end{equation}
or
\begin{equation}
\begin{cases}
h_{x}(k_{c})=\frac{2n-1}{2}\pi, & n\in\mathbb{Z}\\
s'_{c}=\frac{2q-1}{2h_{y}(k_{c})}\in(0,1), & q\in\mathbb{Z}
\end{cases}\label{eq:kcsc22}
\end{equation}
where $s'_{c}=2-s_{c}$. Given the explicit forms of $h_{x}(k)$ and
$h_{y}(k)$, these conditions then determine all possible critical
momenta $k_{c}$ and critical time $t_{c}=s_{c}+2\ell$ ($\ell\in\mathbb{Z}$)
of Floquet DQPTs in the class of periodically quenched systems considered
in this work. Note in passing that the critical time $s_c$ in 
Eqs.~(\ref{eq:kcsc11}) and (\ref{eq:kcsc21}) take universal forms and independent of the 
explicit expressions of $h_x(k)$ and $h_y(k)$, which indicates their connections with the 
global nature of the underlying Floquet states. 
In the meantime, we notice that when the Floquet eigenphase
spectrum of the system becomes gapless, i.e., when $E_{\pm}(k)=0$ or $E_{\pm}(k)=\pm\pi$
in Eq.~(\ref{eq:Ek}), we would obtain 
\begin{equation}
\begin{cases}
h_{x}(k_{0})=m\pi, & m\in\mathbb{Z}\\
h_{y}(k_{0})=n\pi, & n\in\mathbb{Z}
\end{cases}\label{eq:k0}
\end{equation}
where the set $\{k_{0}\}$ gives the locations at which the
eigenphase band-gap closes in $k$-space, which is usually accompanied
by a transition between different Floquet topological phases~\cite{ZhouSQKR}.
From Eqs.~(\ref{eq:kcsc11}) and (\ref{eq:kcsc21}), it is clear that
once $m,n\neq0$ in Eq.~(\ref{eq:k0}), the gapless quasimomenta $\{k_{0}\}$
of the Floquet spectrum are also critical momenta of the Floquet DQPTs,
yielding critical times at $s_{c}$ and $2-s_{c}$ within a driving
period. Since the set $\{k_{0}\}$ is generally different for different
values of $(m,n)$ in Eq.~(\ref{eq:k0}), the possible critical momenta
and critical times in Eqs.~(\ref{eq:kcsc11}) and (\ref{eq:kcsc21})
will also be different in general when the system is initialized in
different Floquet topological phases. This allows one to discriminate
Floquet topological phases from the behaviors of Floquet DQPTs, thus
extending the connection between Floquet DQPTs and Floquet topological
matter found in Ref.~\cite{DQPTExp11} to more general situations. 

\subsection{Rate function of return probability}
We next introduce the rate function of return probability, which is
the observable to show the time-domain nonanalyticity of Floquet
DQPTs explicitly. Following Eq.~(\ref{eq:Gvkt}), the return probability
of a Floquet state $|\psi_{v}(k)\rangle$ ($v=\pm$) evolved from
time zero to $t$ is given by
\begin{equation}
g_{v}(k,t)=|\langle\psi_{v}(k)|U(k,t)|\psi_{v}(k)\rangle|^{2}=g_{v}(k,s),\label{eq:gvkt}
\end{equation}
where 
\begin{equation}
g_{v}(k,s)=|\langle\psi_{v}(k)|U(k,s)|\psi_{v}(k)\rangle|^{2},\label{eq:gvks}
\end{equation}
and $t=s+2\ell$ for $s\in[0,2)$, $\ell\in\mathbb{Z}$. It is clear
that $g_{v}(k,t)$ possesses the same temporal periodicity as the
system Hamiltonian, i.e., $g_{v}(k,t)=g_{v}(k,t+2)$. The complete
behavior of $g_{v}(k,t)$ at any time $t$ is then determined once
$g_{v}(k,s)$ is obtained. For our periodically quenched system, $g_{v}(k,s)$
takes the following explicit form according to Eq.~(\ref{eq:Gvks2})
\begin{equation}
g_{v}(k,s)=\begin{cases}
\cos^{2}[sh_{x}(k)]+\sin^{2}[sh_{x}(k)]n_{x}^{2}(k), & s\in[0,1)\\
\cos^{2}[s'h_{y}(k)]+\sin^{2}[s'h_{y}(k)]n_{y}^{2}(k), & s\in[1,2)
\end{cases}\label{eq:gvks2}
\end{equation}
where $s'=2-s\in[0,1)$. Again, we notice that $g_{v}(k,s)$ is independent
of $v$~($=\pm$), i.e., in which band the system is initialized, and $g_{v}(k,s)=0$
at each given set of critical momentum and time $(k,s)=(k_{c},s_{c})$ in Eqs.~(\ref{eq:kcsc11})-(\ref{eq:kcsc22}).
The latter suggests us to consider the rate function of return probability
$f_{v}(t)$, defined as
\begin{equation}
f(t)=-\lim_{N\rightarrow\infty}\frac{1}{N}\sum_{k}\ln[g(k,t)]=-\int_{-\pi}^{\pi}\frac{dk}{2\pi}\ln[g(k,t)],\label{eq:ft}
\end{equation}
where $N$ is the number of degrees of freedom in the system (e.g.,
the number of unit cells), which is taken to be infinity in the thermodynamic
limit. We have also denoted $g_{v}(k,s)$ by $g(k,s)$ in Eq.~(\ref{eq:ft}), as the former
is independent of $v$. The rate function thus defined may then be viewed as a ``free energy density'' of the Floquet dynamics.
Since $f(t)$ is a periodic function of time
with the period $T=2$, i.e., $f(t)=f(t+2)$, we have for any time
$t=s+2\ell$ with $s\in[0,2)$ and $\ell\in\mathbb{Z}$ that
\begin{equation}
f(t)=f(s)=-\int_{-\pi}^{\pi}\frac{dk}{2\pi}\ln[g(k,s)].\label{eq:fs}
\end{equation}
If $g(k,s)$ vanishes at certain critical time and momentum $(k_{c},t_{c})$,
$\ln[g(k_{c},s_{c})]$ will diverge, and after the integration over
$k$, $f(s)$ could become nonanalytic at time $t_{c}=s_{c}+2\ell$
for all $\ell\in\mathbb{Z}$, which are referred to as Floquet DQPTs.
For 1D systems, a Floquet DQPT is usually manifested
as a cusp in the profile of $f(t)$ around $t=t_{c}$~\cite{DQPTExp11}. Using Eqs.~(\ref{eq:gvks2}) and 
(\ref{eq:fs}), we can further obtain the explicit
expression of the return rate function $f(s)$ for our periodically
quenched system, i.e., 
\begin{equation}
f(s)=-\int_{-\pi}^{\pi}\frac{dk}{2\pi}\cdot\begin{cases}
\ln\{\cos^{2}[sh_{x}(k)]+\sin^{2}[sh_{x}(k)]n_{x}^{2}(k)\},\\
\ln\{\cos^{2}[s'h_{y}(k)]+\sin^{2}[s'h_{y}(k)]n_{y}^{2}(k)\},
\end{cases}\label{eq:fs2}
\end{equation}
for $s\in[0,1)$ and $s'=2-s\in[0,1)$, respectively. The profile of $f(s)$ over a driving period $s\in[0,2)$
then contains all the information about the time-domain nonanalytic
properties of Floquet DQPTs.

\subsection{Geometric phase and DTOP}
In previous studies, dynamical topological order parameters (DTOPs)
have been introduced to characterize DQPTs following a sudden quench~\cite{DTOP1,DTOP2}, or Floquet DQPTs induced by harmonic driving fields~\cite{DQPTExp11}.
The values of a DTOP, obtained from the winding number of the noncyclic
(Pancharatnam) geometric phase of the evolving state over the first
Brillouin zone at each time, usually show a quantized jump at every
critical time of a DQPT or Floquet DQPT, and remain quantized otherwise.
It thus plays the role of a topological order parameter of the dynamics.
In this work, we extend the DTOP-characterization of Floquet DQPTs
to our periodically quenched setting. 

The noncyclic geometric phase is given by the difference between the
total phase and dynamical phase of the return amplitude $G_{v}(k,t)$
($v=\pm$). Following Eq.~(\ref{eq:Gvks}), the total phase is
\begin{equation}
\phi_{v}(k,t)=-i\ln\left[\frac{G_{v}(k,t)}{|G_{v}(k,t)|}\right]=-\ell E_{v}(k)+\phi_{v}(k,s),\label{eq:PhiTot}
\end{equation}
where we have set $t=s+2\ell$ for $s\in[0,2)$ and $\ell\in\mathbb{Z}$
to reach the second equality. Using Eq.~(\ref{eq:Gvks2}), we obtain
the micromotion part of the total phase $\phi_{v}(k,s)$ for our periodically
quenched setting as
\begin{equation}
\phi_{v}(k,s)=\begin{cases}
-\arctan\{\tan[sh_{x}(k)]vn_{x}(k)\},\\
\arctan\{\tan[s'h_{y}(k)]vn_{y}(k)\}-E_{v}(k),
\end{cases}\label{eq:PhiTot2}
\end{equation}
for $s\in[0,1)$ and $s'=2-s\in[0,1)$, respectively. The dynamical phase of the return amplitude is defined
as
\begin{equation}
\phi_{v}^{{\rm D}}(k,t)=-\int_{0}^{t}dt'\langle\psi_{v}(k,t')|H(k,t')|\psi_{v}(k,t')\rangle.\label{eq:PhiDyn}
\end{equation}
Using the time-periodicity of Hamiltonian $H(k,t)$ and the direct
result of Floquet theorem $|\psi_{v}(k,t+2)\rangle=e^{-iE_{v}(k)}|\psi_{v}(k,t)\rangle$,
$\phi_{v}^{{\rm D}}(k,t)$ can be further expressed as
\begin{equation}
\phi_{v}^{{\rm D}}(k,t)=\ell\phi_{v}^{{\rm D}}(k,2)+\phi_{v}^{{\rm D}}(k,s),\label{eq:PhiDyn2}
\end{equation}
where we have set $t=s+2\ell$ with $s\in[0,2)$ and $\ell\in\mathbb{Z}$
on the right hand side of the equality. For our periodically quenched
system, we find with the help of Eqs.~(\ref{eq:Hkt}), (\ref{eq:Psivk}), (\ref{eq:Psivkt1}) and (\ref{eq:Psivkt2}) that $\phi_{v}^{{\rm D}}(k,2)=-v\vec{h}(k)\cdot\vec{n}(k)$ and
\begin{equation}
\phi_{v}^{{\rm D}}(k,s)=\begin{cases}
-vsh_{x}(k)n_{x}(k), & s\in[0,1)\\
-v[\vec{h}(k)\cdot\vec{n}(k)-s'h_{y}(k)n_{y}(k)], & s\in[1,2)
\end{cases}\label{eq:PhiDyn3}
\end{equation}
where $s'=2-s$. Note that the dynamical phase is closely related to the
relative angle between the components $[h_{x}(k),h_{y}(k),0]$ of
the time-dependent Hamiltonian $H(k,t)$ and the components $[n_{x}(k),n_{y}(k),n_{z}(k)]$
of the Floquet effective Hamiltonian in each half of the driving period.
Taking the difference between the total phase and dynamical phase,
we find the noncyclic geometric phase~\cite{DQPTExp11} to be
\begin{equation}
\phi_{v}^{{\rm G}}(k,t)=\phi_{v}(k,t)-\phi_{v}^{{\rm D}}(k,t),\label{eq:PhiGeo}
\end{equation}
whose explicit expression for our periodically quenched system can
be obtained directly from Eqs.~(\ref{eq:PhiTot})-(\ref{eq:PhiDyn3}).
At a fixed time $t$, the winding number of $\phi_{v}^{{\rm G}}(k,t)$
in $k$-space then determines the DTOP $w_{v}(t)$, given by
\begin{equation}
w_{v}(t)=\int\frac{dk}{2\pi}\partial_{k}\phi_{v}^{{\rm G}}(k,t).\label{eq:DTOP}
\end{equation}
When the evolution of the system passes a critical time $t_{c}$ of
the Floquet DQPTs, the value of $w_{v}(t)$ is expected to show a quantized
jump~\cite{DQPTExp11}, i.e.,
\begin{equation}
w_{v}(t_{c}^{+})-w_{v}(t_{c}^{-})\in\mathbb{Z},\label{eq:wtpm}
\end{equation}
where the exact amount of jump depends on the configuration of
the critical momenta set $\{k_{c}\}$ in the first Brillouin zone.
For certain specific models, the integration range of $k$ in Eq.~(\ref{eq:wtpm}) 
could be chosen as a subset of the first Brillouin
zone $k\in[-\pi,\pi)$ due to the possible translation and reflection
symmetries of $\phi_{v}^{{\rm G}}(k,t)$ in $k$-space. For example,
if we have $h_{j}(k\pm\pi)=-h_{j}(k)$ for $j=x,y$, both $\phi_{v}(k,t)$
and $\phi_{v}^{{\rm D}}(k,t)$ will be invariant under the translation
over $\pi$ in $k$-space, and the same is true for $\phi_{v}^{{\rm G}}(k,t)$.
So the range of integration in Eq.~(\ref{eq:wtpm}) can be confined
to $k\in[0,\pi]$ in this case. More examples about the reduction
of integration range in the calculation of $w_{v}(t)$ will be made
clear in the next section.

To sum up, we have established a theoretical framework to analyze
Floquet DQPTs in a typical class of periodically quenched lattice
model. The critical conditions as summarized in Eqs.~(\ref{eq:kcsc11})-(\ref{eq:kcsc22})
clearly suggest that by enlarging the amplitudes of quench functions
$h_{x}(k)$ and $h_{y}(k)$, the exact number of Floquet DQPTs that
could occur in each driving periodic can be put under flexible control.
This highlights one of the key advantages of Floquet engineering in
the realization of DQPTs. Besides, we have observed the connection
between Floquet DQPTs and the underlying Floquet topological phases
of the periodically quenched system, and also introduced a DTOP to
characterize the topological signatures of Floquet DQPTs. In the following
section, we will demonstrate our results with a prototypical periodically
quenched lattice model, which is rich in Floquet DQPTs and experimentally
realizable at the same time.

\section{Results}\label{sec:Res}

In this section, we demonstrate our theoretical scheme of Floquet
DQPTs in a piecewise quenched lattice (PQL) model, which possesses
rich Floquet topological phases and also being realizable in quantum
simulator setups.

\subsection{The PQL model}
Our explicit model is obtained by setting 
\begin{equation}
h_{x}(k)=J_{x}\cos k,\quad h_{y}(k)=J_{y}\sin k,\label{eq:hxyk}
\end{equation}
in the general Hamiltonian Eq.~(\ref{eq:Hkt}), leading to the Floquet
operator
\begin{equation}
U(k)=e^{-iJ_{y}\sin k\sigma_{y}}e^{-iJ_{x}\cos k\sigma_{x}}.\label{eq:UkSKR}
\end{equation}
Here the parameters $J_{x}$ and $J_{y}$ control the amplitudes of
the two quenches within each half of the driving period. Physically,
they can be interpreted as the kicking strengths of a spin-$1/2$
quantum kicked rotor~\cite{ZhouSQKR}, or the hopping amplitudes of a particle
on a tight-binding lattice~\cite{ZhouDWN}. In the former case, the system
may be realized by the setup proposed in Refs.~\cite{SQKRExp1,SQKRExp2}, which achieves
the quantum walk in momentum space by $^{87}$Rb BEC atoms. In the
latter case, the system may be realized by replacing the harmonic
driving fields with square-wave pulses in the NV center experimental
setup of Refs.~\cite{DQPTExp11,ZhouGTPExp}. Theoretically, rich Floquet topological phases
have been identified in this model, which are characterized by larger
topological invariants and multiple topological edge modes~\cite{ZhouSQKR}.
This further motivates us to choose this model as our working example
to demonstrate the key features of Floquet DQPTs in periodically quenched
systems.

From now on, we take the quench amplitudes $J_{x,y}>0$ without loss
of generality. Using Eqs.~(\ref{eq:hxyk}) and (\ref{eq:UkSKR}),
the Floquet operator can be written in a single exponential form as
$U(k)=e^{-iE(k)\vec{n}(k)\cdot\vec{\sigma}}$, where $\cos[E(k)]=\cos(J_{x}\cos k)\cos(J_{y}\sin k)$
and the unit vector $\vec{n}(k)=[n_{x}(k),n_{y}(k),n_{z}(k)]$, with
\begin{alignat}{1}
n_{x}(k) & =\frac{\sin(J_{x}\cos k)\cos(J_{y}\sin k)}{\sin[E(k)]},\nonumber \\
n_{y}(k) & =\frac{\sin(J_{y}\sin k)\cos(J_{x}\cos k)}{\sin[E(k)]},\label{eq:nxyzk}\\
n_{z}(k) & =-\frac{\sin(J_{x}\cos k)\sin(J_{y}\sin k)}{\sin[E(k)]}.\nonumber 
\end{alignat}
The Floquet effective Hamiltonian then takes the form $H_{{\rm eff}}=E(k)\vec{n}(k)\cdot\vec{\sigma}=\vec{d}(k)\cdot\vec{\sigma}$,
with the corresponding eigenvalue equation $H_{{\rm eff}}(k)|\psi_{\pm}(k)\rangle=E_{\pm}(k)|\psi_{\pm}(k)\rangle$,
whose explicit solutions are given by Eq.~(\ref{eq:Psivk}). Following
Eq.~(\ref{eq:k0}), the gapless quasimomenta of the model satisfy
$J_{x}\cos k_{0}=m\pi$ and $J_{y}\sin k_{0}=n\pi$ for $m,n\in\mathbb{Z}$,
yielding the curves that describing the boundaries between different
Floquet topological phases, i.e., $m^{2}\pi^{2}/J_{x}^{2}+n^{2}\pi^{2}/J_{y}^{2}=1$
for $|m|\leq J_{x}/\pi$ and $|n|\leq J_{y}/\pi$~\cite{ZhouSQKR}. According
to the discussion in the last section, the solutions of $k_{0}$ also
generate a set of critical momenta $k_{c}$ for the Floquet DQPTs
when $m,n\neq0$.

Using Eqs.~(\ref{eq:kcsc11})-(\ref{eq:kcsc22}), we could obtain
the solutions of critical momenta and critical times for the PQL model if the
system is initialized in a Floquet eigenstate. In the first half of
a driving period, i.e., for $s\in[0,1)$, the set of $(k_{c},s_{c})$ satisfy
\begin{equation}
\begin{cases}
k_{c}=\arccos\left(\frac{m\pi}{J_{x}}\right), & m\in\mathbb{Z}_{\neq0}\cap|m|\leq\frac{J_{x}}{\pi}\\
s_{c}=\frac{2p-1}{2m}\in[0,1), & p\in\mathbb{Z}
\end{cases}\label{eq:PQLksc11}
\end{equation}
or
\begin{equation}
\begin{cases}
k_{c}=\arcsin\left(\frac{2m-1}{2J_{y}}\pi\right), & m\in\mathbb{Z}\cap\left|m-\frac{1}{2}\right|\leq\frac{J_{y}}{\pi}\\
s_{c}=\frac{2p-1}{2J_{x}\cos k_{c}}\pi\in[0,1). & p\in\mathbb{Z}
\end{cases}\label{eq:PQLksc12}
\end{equation}
Similarly, in the second half of a driving period, i.e., for $s\in[1,2)$,
the set $(k_{c},s_{c}=2-s'_{c})$ satisfy
\begin{equation}
\begin{cases}
k_{c}=\arcsin\left(\frac{n\pi}{J_{y}}\right), & n\in\mathbb{Z}_{\neq0}\cap|n|\leq\frac{J_{y}}{\pi}\\
s'_{c}=\frac{2q-1}{2n}\in[0,1), & q\in\mathbb{Z}
\end{cases}\label{eq:PQLksc21}
\end{equation}
or
\begin{equation}
\begin{cases}
k_{c}=\arccos\left(\frac{2n-1}{2J_{x}}\pi\right), & n\in\mathbb{Z}\cap\left|n-\frac{1}{2}\right|\leq\frac{J_{x}}{\pi}\\
s'_{c}=\frac{2q-1}{2J_{y}\sin k_{c}}\pi\in[0,1). & q\in\mathbb{Z}
\end{cases}\label{eq:PQLksc22}
\end{equation}
Eqs.~(\ref{eq:PQLksc11})-(\ref{eq:PQLksc22}) provide us with analytical
solutions for all possible critical momenta $k_{c}$ and critical
times $t_{c}=s_{c}+2\ell$ ($\ell\in\mathbb{Z}$) of the Floquet DQPTs
in the PQL model. The notable connection between the expressions of
$k_{c}$ in Eqs.~(\ref{eq:PQLksc11}), (\ref{eq:PQLksc21}) and the
gapless quasimomentum $k_{0}$ satisfying $k_{0}=\arccos(m\pi/J_{x})=\arcsin(n\pi/J_{y})$
for $m,n\in\mathbb{Z}$ also allows one to obtain qualitatively different
patterns of Floquet DQPTs in different Floquet topological phases
of the PQL model, as will be demonstrated explicitly by numerical
calculations.

\subsection{Numerical results}
In the following, we present typical examples of Floquet DQPTs in
the PQL model by evaluating the rate function of return probability
in Eq.~(\ref{eq:ft}), noncyclic geometric phase in Eq.~(\ref{eq:PhiGeo})
and DTOP in Eq.~(\ref{eq:DTOP}). 
We first evolve the system according to Eq.~(\ref{eq:gvkt}) for each Floquet eigenstate, and then obtain the return rate from Eq.~(\ref{eq:ft}) by numerically evaluating the integral over quasimomentum $k$. The geometric phase is obtained by taking the difference between the total phase and the dynamical phase of the return amplitude following Eqs.~(\ref{eq:PhiTot}) and (\ref{eq:PhiDyn}). After numerically obtaining the geometric phase, we compute the DTOP in terms of Eq.~(\ref{eq:DTOP}). In the computation of the integral, $N=300$ grids in $k$-space is taken, which is enough for the convergence of the numerical data in order to reproduce the nonanalytic signals of Floquet DQPTs in the thermodynamic limit (corresponding to $N\rightarrow\infty$).
Note that for the PQL model, we
have $h_{j}(k\pm\pi)=-h_{j}(k)$ for $j=x,y$ according to Eq.~(\ref{eq:hxyk}),
implying that the geometric phase has a translational symmetry over
$\pi$ in $k$-space. Furthermore, since $h_{x}(k)$ {[}$h_{y}(k)${]}
and $n_{x}(k)$ {[}$n_{y}(k)${]} are even (odd) functions of $k$,
the geometric phase also has a reflection symmetry with respect to
$k=0$ according to Eqs.~(\ref{eq:PhiTot2}), (\ref{eq:PhiDyn3})
and (\ref{eq:PhiGeo}). Therefore, in the calculation of DTOP in Eq.~(\ref{eq:DTOP}) 
for the PQL model, we could restrict the range of
integration over $k$ to one quarter of the first Brillouin zone,
e.g., $k\in[0,\pi/2]$. As a further simplification, we notice that
the order of the two quenches are swapped under the combined effects
of parameter exchange $J_{x}\leftrightarrow J_{y}$ and momentum shift $k\rightarrow\pi/2-k$.
The result of such a swap is simply a shifting of the Floquet DQPTs
from one half of the driving period to the other half, as directly
deducible from Eq.~(\ref{eq:fs2}). Therefore, for $J\neq J'$, the
Floquet DQPTs in the system with $(J_{x},J_{y})=(J,J')$ and $(J_{x},J_{y})=(J',J)$
can be directly related with each other. For $J=J'$, each Floquet DQPT happens at the micromotion time $s_c$ has a mirror that happens at $2-s_c$ for all $t_c=s_c+2\ell$ with $\ell\in{\mathbb Z}$. 

\begin{figure}
	\begin{centering}
		\includegraphics[scale=0.5]{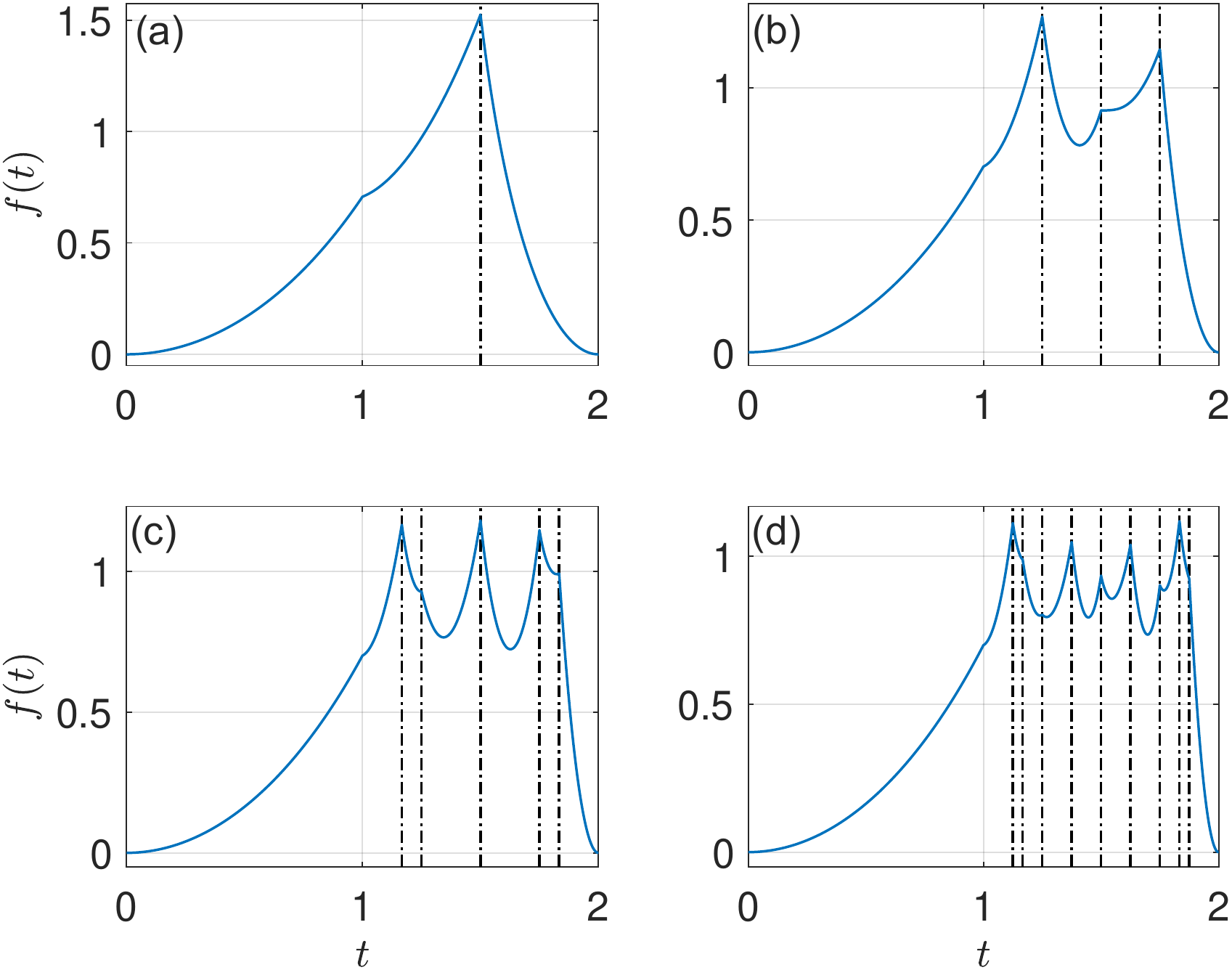}
		\par\end{centering}
	\caption{Rate function $f(t)$ of the PQL model (blue solid lines) over a driving
		period $t\in[0,2]$, in which a Floquet DQPT manifests as a cusp at
		the corresponding critical time. System parameters are set as $J_{x}=0.5\pi$
		for all panels and $J_{y}=1.1\pi,2.1\pi,3.1\pi,4.1\pi$ for panels
		(a), (b), (c) and (d), respectively. The crossing points of black dash-dotted lines
		with the horizontal axis in each panel refer to the critical times
		$t_{c}$ of Floquet DQPTs in each case, which are obtained analytically
		from Eq.~(\ref{eq:PQLksc21}).\label{fig:RF1}}
\end{figure}

\begin{figure}
	\begin{centering}
		\includegraphics[scale=0.49]{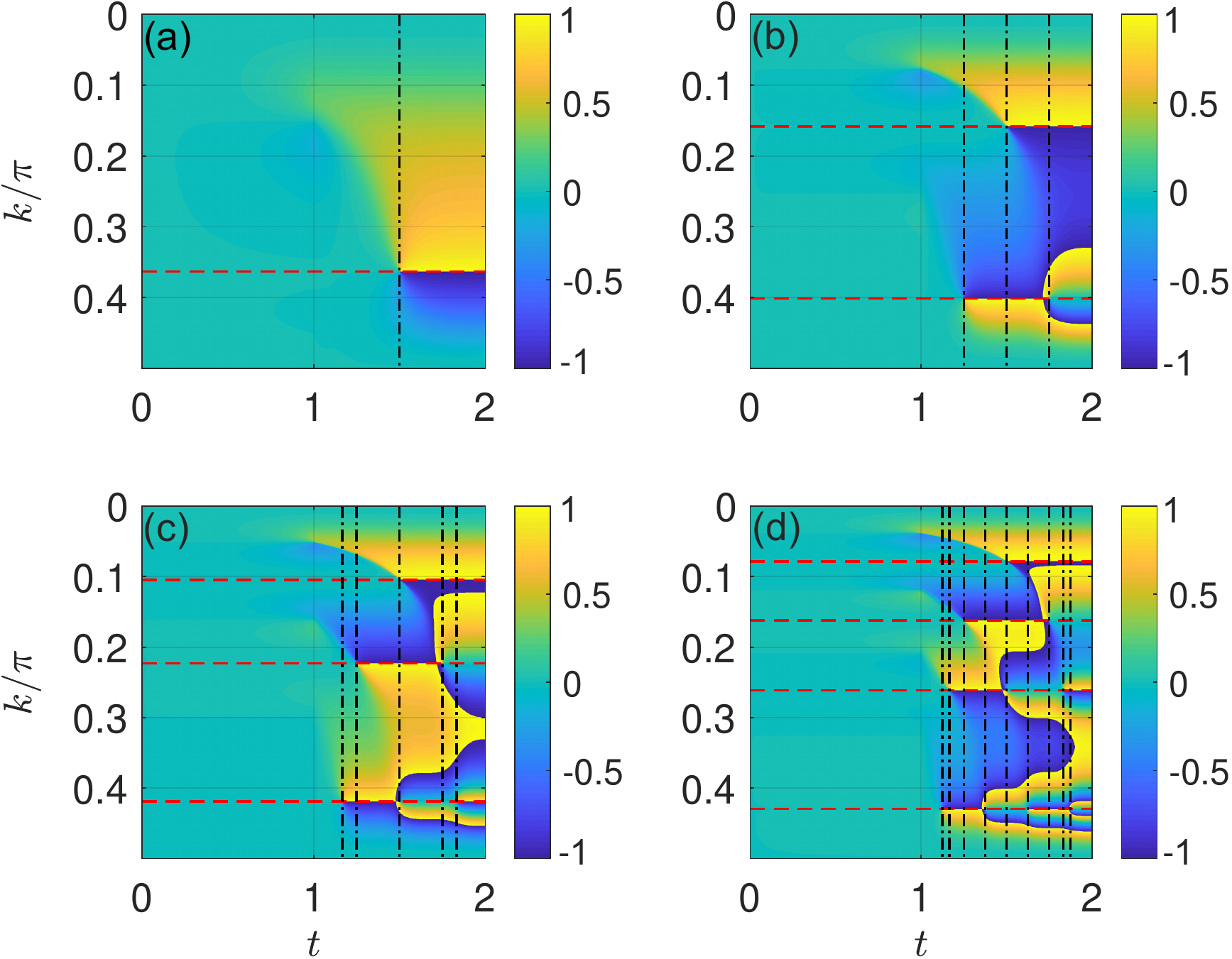}
		\par\end{centering}
	\caption{Noncyclic geometric phase $\phi_{-}^{{\rm G}}(k,t)/\pi$ of the PQL
		model versus the quasimomentum $k\in[0,\pi/2]$ and time $t\in[0,2]$.
		System parameters are set as $J_{x}=0.5\pi$ for all panels and $J_{y}=1.1\pi,2.1\pi,3.1\pi,4.1\pi$
		for panels (a), (b), (c) and (d), respectively. In each panel, the crossing points
		of the horizontal red dashed lines with the vertical axis correspond
		to the critical momenta $k_{c}$, and the intersection points of the
		vertical black dash-dotted lines with the horizontal axis refer to
		the critical times $t_{c}$, which are obtained from Eq.~(\ref{eq:PQLksc21})
		analytically for each chosen set of system parameters.\label{fig:GP1}}
\end{figure}

\begin{figure}
	\begin{centering}
		\includegraphics[scale=0.5]{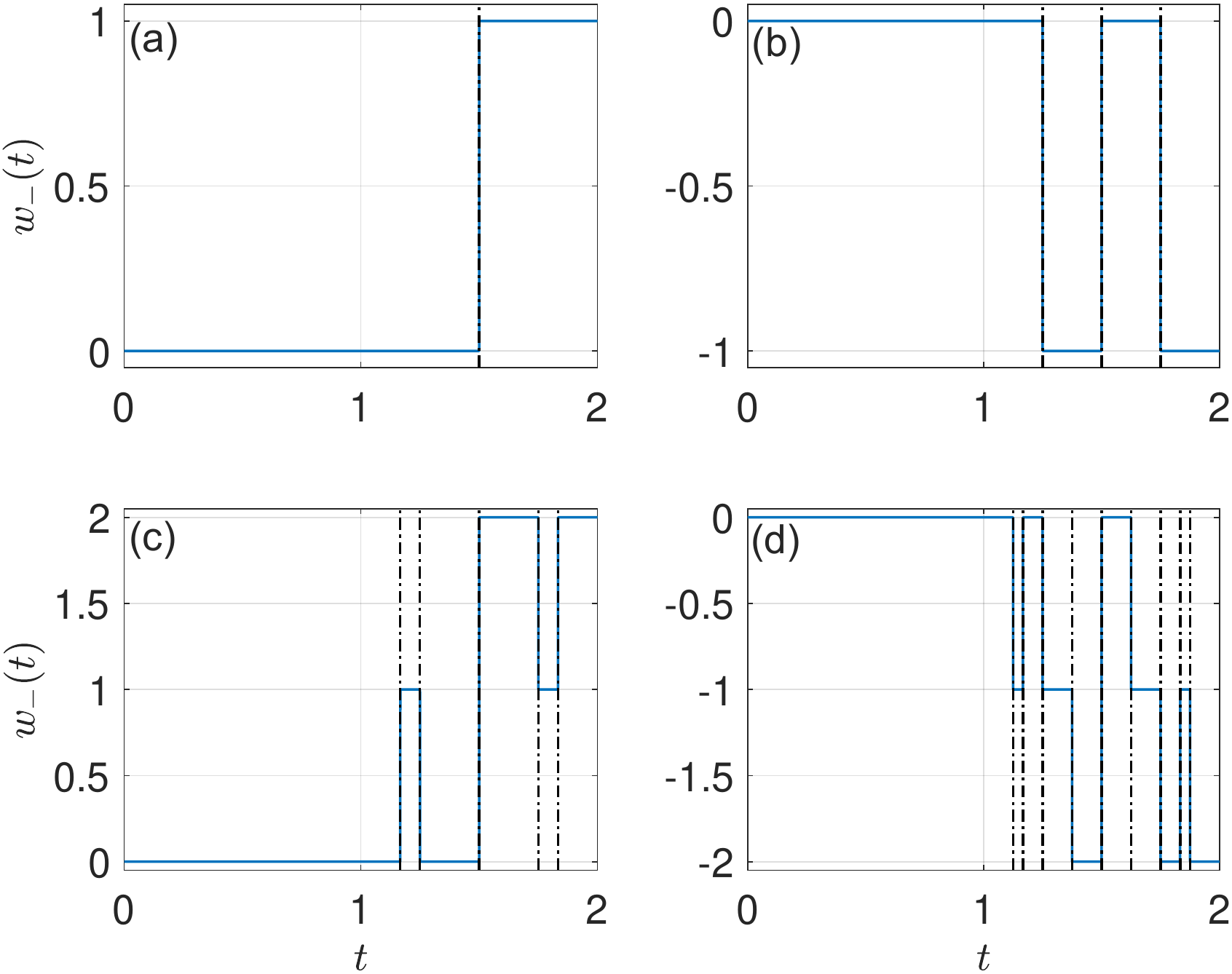}
		\par\end{centering}
	\caption{DTOP $w_{-}(t)$ of the PQL model (blue solid lines) versus time over
		a driving period $t\in[0,2]$. System parameters are chosen to be
		$J_{x}=0.5\pi$ for all panels and $J_{y}=1.1\pi,2.1\pi,3.1\pi,4.1\pi$
		for panels (a), (b), (c) and (d). In each panel, the crossing points
		of the vertical black dash-dotted lines with the horizontal axis refer
		to the critical times $t_{c}$, which are obtained analytically with
		the chosen system parameters from Eq.~(\ref{eq:PQLksc21}).
		The DTOP makes a quantized jump at each critical time $t_{c}$ and remaining quantized otherwise.\label{fig:DTOP1}}
\end{figure}

In the first set of results, we consider the initial state of the
PQL model to be prepared in the Floquet band with quasienergy $E_{-}(k)$
at each quasimomentum $k\in[-\pi,\pi)$, and present the rate function of return
probability $f(t)$ {[}Eq.~(\ref{eq:fs2}){]}, noncyclic geometric
phase $\phi_{-}^{{\rm G}}(k,t)$ {[}Eq.~(\ref{eq:PhiGeo}){]} and
DTOP $w_{-}(t)$ {[}Eq.~(\ref{eq:DTOP}){]} versus time over a driving
period $t=0\rightarrow2$ in Figs.~\ref{fig:RF1}, \ref{fig:GP1}
and \ref{fig:DTOP1}, respectively. The system parameters are set
as $J_{x}=0.5\pi$ for all figure panels and $J_{y}=1.1\pi,2.1\pi,3.1\pi,4.1\pi$
in the panels (a), (b), (c) and (d) of each figure, respectively.
According to Ref.~\cite{ZhouSQKR}, the PQL model with these four sets of system
parameters belong to four different Floquet topological phases, with
larger topological invariants at larger values of $J_{y}$. In Fig.~\ref{fig:RF1}, 
we observe that with the increase of $J_{y}$, the
system undergoes more and more Floquet DQPTs within a single driving
period, with a cusp observed at every critical time $t_{c}$ of the
DQPT. The possible values of critical times are predicted precisely
by the critical condition in Eq.~(\ref{eq:PQLksc21}), whose solutions
of $k_{c}$ are closely related to the gapless quasimomenta $k_{0}$
of the system. Specially, more Floquet DQPTs are observed when the
initial phase of the PQL model carries larger topological invariants~(see the phase diagram in Fig.~1 of Ref.~\cite{ZhouSQKR}).
This implies the possibility of discriminating different Floquet topological
phases of the PQL model from Floquet DQPTs, and also introduces a
route to engineer more Floquet DQPTs in a finite time window with
the help of Floquet topological phases. Besides, the global profiles
of $f(t)$ in Fig.~\ref{fig:RF1} do not decay with time thanks to
the applied driving field, which allow us to observe and detect the
nonanalytic signatures of Floquet DQPTs in a larger time window compared
with the case in conventional DQPTs following a single quench~\cite{DQPTRev1}. The noncyclic geometric
phase $\phi_{-}^{{\rm G}}(k,t)$, as presented in Fig.~\ref{fig:GP1} also
possesses a $2\pi$-jump around the critical momenta $k_{c}$ (red
dashed lines) at each critical time $t_{c}$ (black dash-dotted lines),
reflecting the discontinuity in the time-derivatives of the rate function
$f(t)$ at the Floquet DQPTs. After the integration over $k\in[0,\pi/2]$,
the $2\pi$-change of $\phi_{-}^{{\rm G}}(k,t)$~(corresponding to the sudden change of colors from dark blue to dark yellow or the opposite in each figure panel)
finally leads to
the integer quantized jump of DTOP at each critical time (black dash-dotted
lines) of the Floquet DQPT, as presented in Fig.~\ref{fig:DTOP1}.
Except at the critical times, the DTOP remain quantized over other
time durations, and thus playing the role of a topological order parameter
for Floquet DQPTs. Notably, in Figs.~\ref{fig:DTOP1}(c) and \ref{fig:DTOP1}(d),
$w_{-}(t)$ possesses a change $\Delta{w_-}=2$ at $t_{c}=1.5$. Such a jump
of DTOP in a step bigger than $1$ is not observed in DQPTs
following a single quantum quench, and may thus be a unique feature
of DQPTs following time-periodic drivings or other more complicated
nonequilibrium protocols.

\begin{figure}
	\begin{centering}
		\includegraphics[scale=0.5]{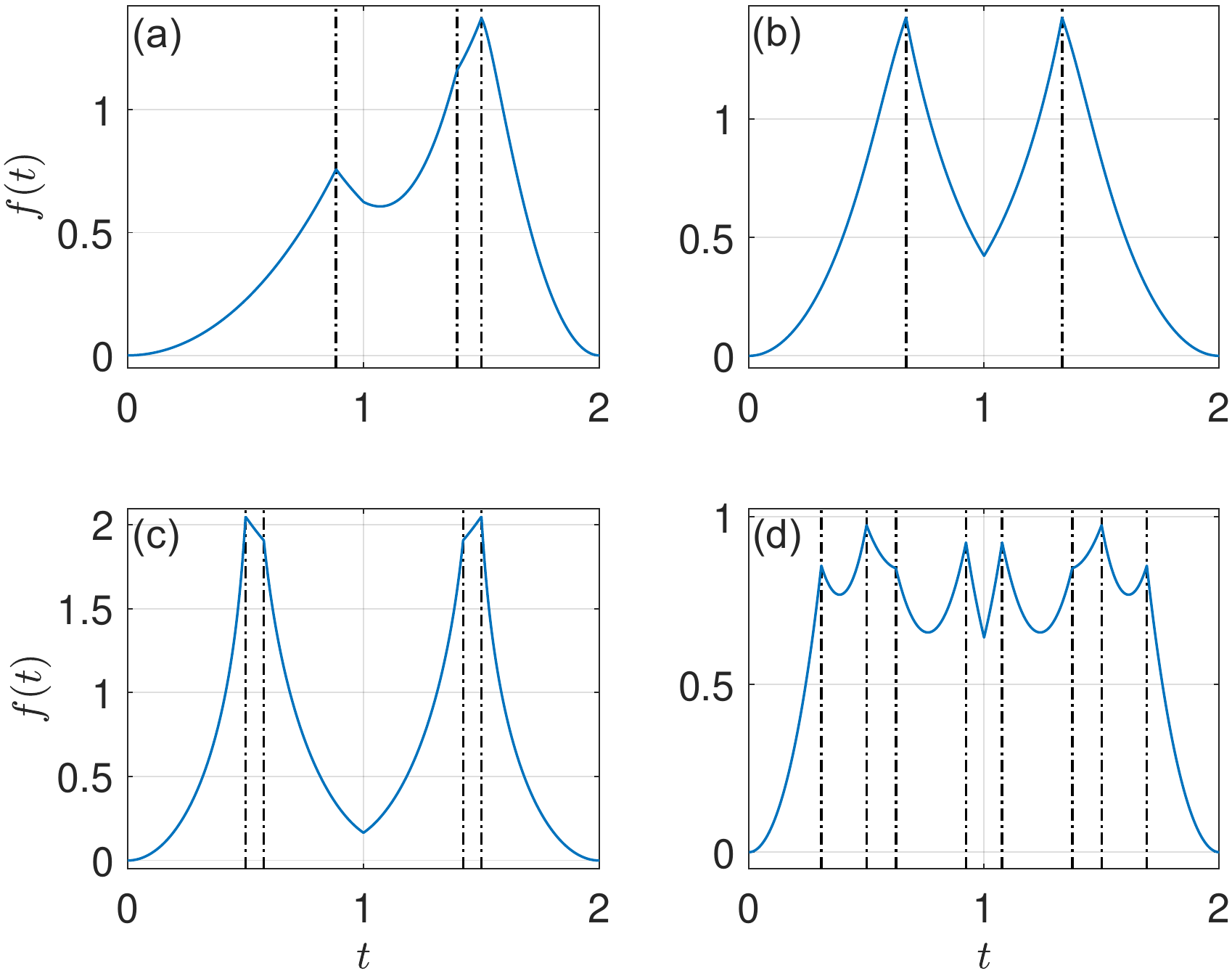}
		\par\end{centering}
	\caption{Rate function $f(t)$ of the PQL model (blue solid lines) versus time
		over a driving period, in which a Floquet DQPT manifests as a cusp
		at the corresponding critical time. System parameters are set as $(J_{x},J_{y})=(0.6\pi,1.5\pi)$
		for panel (a), $(J_{x},J_{y})=(0.9\pi,0.9\pi)$ for panel (b), $(J_{x},J_{y})=(\pi,\pi)$
		for panel (c), and $(J_{x},J_{y})=(1.7\pi,1.7\pi)$ for panel (d).
		The intersection points of black dash-dotted lines with the horizontal
		axis in each panel give the critical times $t_{c}$ of Floquet DQPTs
		in each case, which are obtained analytically from Eqs.~(\ref{eq:PQLksc11})-(\ref{eq:PQLksc22}).\label{fig:RF2}}
\end{figure}

\begin{figure}
	\begin{centering}
		\includegraphics[scale=0.49]{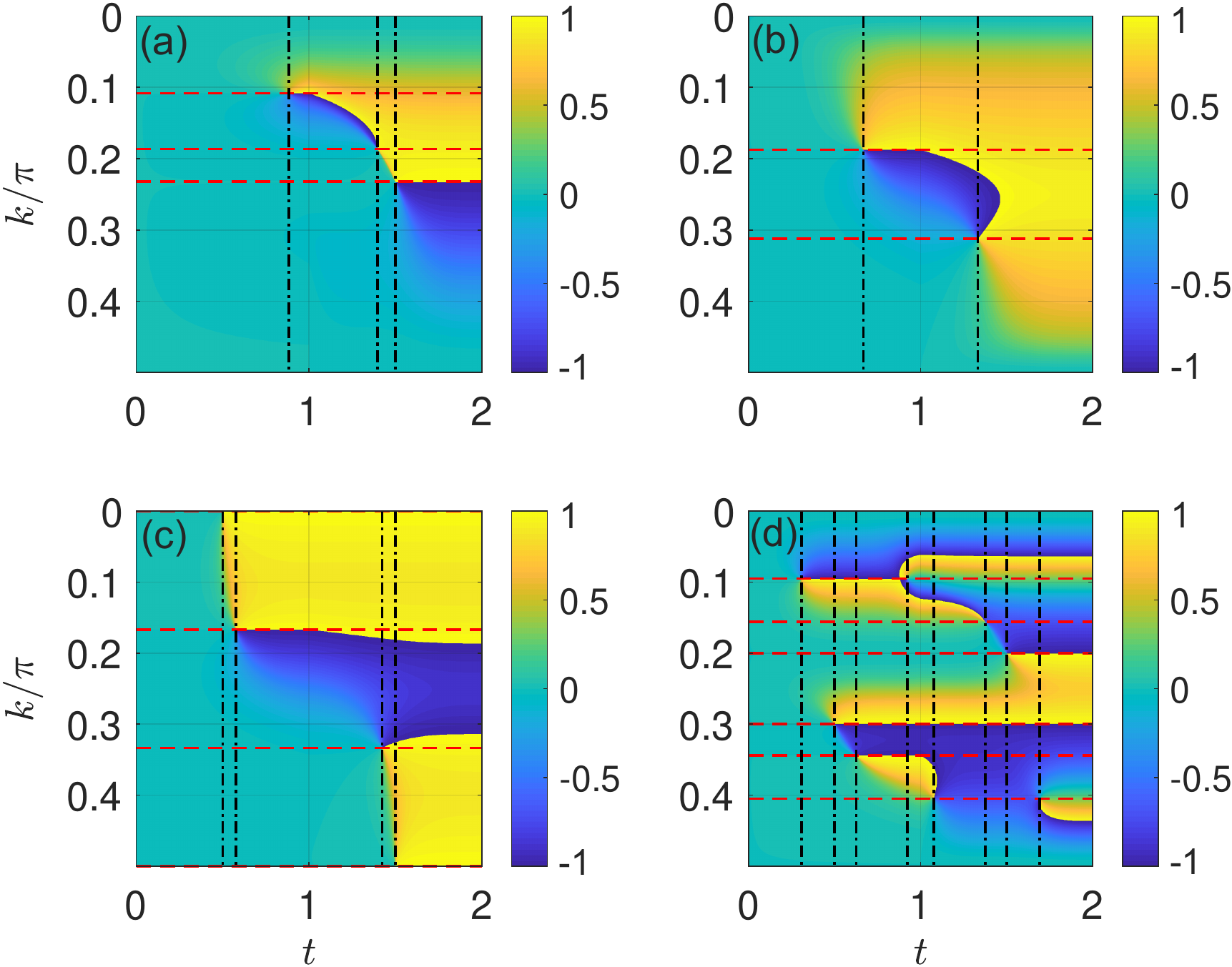}
		\par\end{centering}
	\caption{Noncyclic geometric phase $\phi_{-}^{{\rm G}}(k,t)/\pi$ of the PQL
		model with respect to the quasimomentum $k\in[0,\pi/2]$ and time
		$t\in[0,2]$. System parameters are set as $(J_{x},J_{y})=(0.6\pi,1.5\pi)$
		for panel (a), $(J_{x},J_{y})=(0.9\pi,0.9\pi)$ for panel (b), $(J_{x},J_{y})=(\pi,\pi)$
		for panel (c), and $(J_{x},J_{y})=(1.7\pi,1.7\pi)$ for panel (d).
		In each panel, the intersection points of the horizontal red dashed
		lines with the vertical axis refer to the critical momenta $k_{c}$,
		and the crossing points of the vertical black dash-dotted lines with
		the horizontal axis correspond to the critical times $t_{c}$, both
		of which are obtained from Eqs.~(\ref{eq:PQLksc11})-(\ref{eq:PQLksc22})
		analytically for each chosen set of system parameters.\label{fig:GP2}}
\end{figure}

\begin{figure}
	\begin{centering}
		\includegraphics[scale=0.5]{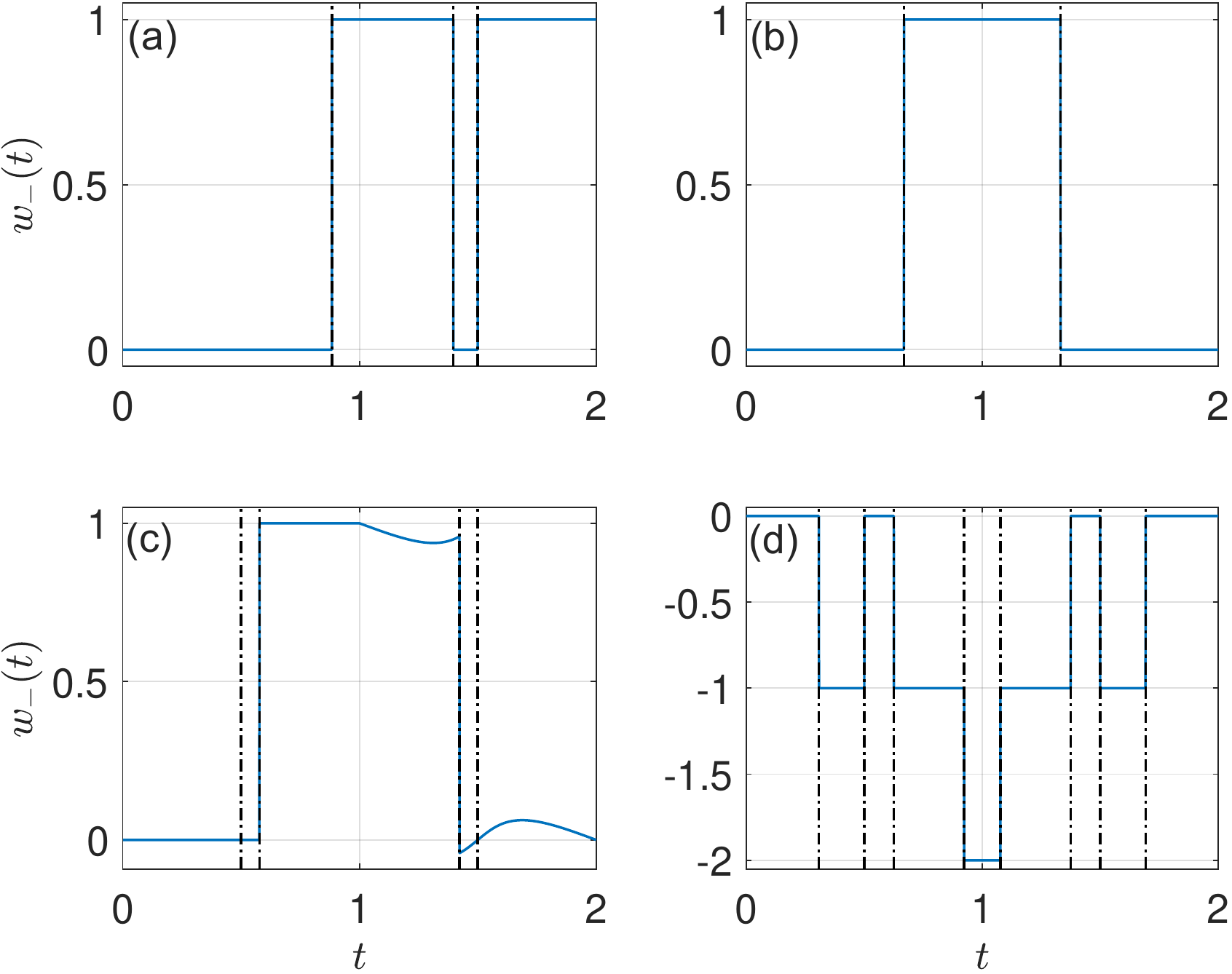}
		\par\end{centering}
	\caption{DTOP $w_{-}(t)$ of the PQL model (blue solid lines) versus time over
		a driving period $t\in[0,2]$. System parameters are set as $(J_{x},J_{y})=(0.6\pi,1.5\pi)$
		for panel (a), $(J_{x},J_{y})=(0.9\pi,0.9\pi)$ for panel (b), $(J_{x},J_{y})=(\pi,\pi)$
		for panel (c), and $(J_{x},J_{y})=(1.7\pi,1.7\pi)$ for panel (d).
		In each panel, the intersection points of the vertical black dash-dotted
		lines with the horizontal axis correspond to the critical times $t_{c}$,
		which are obtained analytically from Eqs.~(\ref{eq:PQLksc11})-(\ref{eq:PQLksc22})
		with the chosen system parameters. The DTOP makes a quantized jump
		at each $t_{c}$ except in panel (c), where the system is initialized
		in a gapless Floquet phase.\label{fig:DTOP2}}
\end{figure}

In the second set of results, we still prepare the initial state of
the system in the Floquet band of the PQL model with negative eigenphases
at all $k\in[-\pi,\pi)$, and investigate the Floquet DQPTs following
quenches over a complete driving period. The quench amplitudes are
chosen as $(J_{x},J_{y})=(0.6\pi,1.5\pi)$, $(0.9\pi,0.9\pi)$, $(\pi,\pi)$
and $(1.7\pi,1.7\pi)$ for the results presented in panels (a), (b),
(c) and (d), respectively, of the Figs.~\ref{fig:RF2}, \ref{fig:GP2}
and \ref{fig:DTOP2}. In panels (a)-(d) of Fig.~\ref{fig:RF2}, we
observe Floquet DQPTs within each half of the driving period, with
the critical times $t_{c}$ (black dash-dotted lines) determined analytically
by Eqs.~(\ref{eq:PQLksc11})-(\ref{eq:PQLksc22}). Notably, when $J_{x}=J_{y}$,
we found the critical times $t_{c}$ to be symmetric with respect
to $t=T/2=1$, which is originated from the symmetry of the rate function
$f(t)$ under the exchange of quenching amplitudes $J_{x}$ and $J_{y}$,
as discussed before in this subsection. Besides, we also observe the
general tendency of having more Floquet DQPTs under larger quenching
amplitudes, which again highlights the flexibility of the Floquet
approach in the engineering and control of DQPTs within a finite
time domain. In Fig.~\ref{fig:GP2}, we observe that the noncyclic
geometric phase $\phi_{-}^{{\rm G}}(k,t)$ develops a $2\pi$-jump
around $k=k_{c}$ (red dashed lines) every time when the system evolves
across a critical time $t_{c}$ (black dash-dotted lines) in each
figure panel~(corresponding to the sudden change of colors from dark blue to dark yellow or the opposite), which provide geometric signatures of the nonanalyticity
for Floquet DQPTs. Finally, in Fig.~\ref{fig:DTOP2}, we present the
DTOP $w_{-}(t)$ versus time over a driving period, as obtained from
Eq.~(\ref{eq:DTOP}). In Figs.~\ref{fig:DTOP2}(a), (b) and (d), the
PQL system is initialized in different Floquet topological phases~\cite{ZhouSQKR}, 
and the qualitatively different behaviors of $w_{-}(t)$ again
allow us to distinguish different Floquet topological phases with
the help of Floquet DQPTs. Interestingly, we observe anomalous behaviors
in the DTOP in Fig.~\ref{fig:DTOP2}(c), in the sense that it either
changes smoothly in certain time windows~(but still with quantized jumps at the critical times), or shows no signatures of jump across certain
critical times. The underlying reason behind such an anomaly is that
for the case considered in Fig.~\ref{fig:DTOP2}(c), the system is
initialized in a gapless Floquet phase, instead of a gapped ``ground
state'' usually employed in the study of DQPTs. Therefore, the Floquet
DQPTs could also provide us with a useful dynamical tool to locate
the phase boundaries between different Floquet topological matter.
For completeness, we have also explored the Floquet DQPTs in other parameter
domains of the PQL model, and obtained results that are consistent with
the general principles as presented in this section, thus verifying the theoretical framework introduced in Sec.~\ref{sec:The} of this work.
As a final comment, for initial states prepared in topologically trivial Floquet phases, no signatures of Floquet DQPTs have been reported in the previous study~\cite{DQPTExp11}. However, this does not exclude the possibility of observing Floquet DQPTs in systems prepared at trivial initial states, as also notified before in the study of non-Floquet DQPTs following a single quench~\cite{DTOP2}. Therefore, the most general relationship between the topological nature of Floquet states and Floquet DQPTs deserves more investigations, which will be left for future studies.

\section{Summary}\label{sec:Sum}
In this work, we introduced a framework to describe Floquet DQPTs in
periodically quenched 1D systems with chiral symmetry. Within a driving
period, we observed a series of Floquet DQPTs, which are characterized
by cusps in the rate function of return probability with a non-decaying global profile, and quantized jumps
of the dynamical topological order parameter~(DTOP). These nonanalytic
features all repeat periodically in time with the same period as the
driving field. Via tuning the strengths of quenches, both the critical
momenta, critical times and number of Floquet DQPTs within a driving
period can be put under flexible control, revealing the advantage
of Floquet engineering in the realization of DQPTs. Furthermore, the
DTOP behaves differently when the underlying Floquet system is prepared
in a gapped topological phase or in a gapless phase, and can thus
be used as an efficient probe to the topological phase transitions
in Floquet systems. Our theoretical scheme is further verified in a prototypical
piecewise quenched lattice model, which holds rich Floquet topological
phases and phase transitions~\cite{ZhouSQKR}. 

By changing the driving field
realized in Ref.~\cite{DQPTExp11} from the harmonic form to the square-wave form,
we expect our predicted signatures of Floquet DQPTs to be readily
detectable in the setup containing an NV center in diamond~\cite{DQPTExp11,ZhouGTPExp}.
In the experiment reported in Ref.~\cite{DQPTExp11}, the finite coherence time of electron spin and the imperfection of microwave pulses may cause uncertainties. The nearly vanishing return probabilities around the critical points of Floquet DQPTs could also introduce some errors in the measurement of rate function at the critical times. However, the main features of Floquet DQPTs in connection with cusps in the rate function and jumps in the DTOP could still be clearly seen over a time window of three driving periods ($6\mu$s), as shown in Ref.~\cite{DQPTExp11}. Such a time window is long enough for one to observe Floquet DQPTs in periodically quenched systems, and the lifetime of Floquet states is also comparable to the observed time of Floquet DQPTs in the experiment~\cite{DQPTExp11}. Very recently, the periodically quenched lattice model introduced in Eq.~(\ref{eq:UkSKR}) have been simulated experimentally, and its Floquet topological phases have also been observed in Ref.~\cite{MFarXiv2020} with an NV center setup by measuring the time-averaged spin textures. Specially, the coupling strengths $(J_x,J_y)$ can reach $3\pi$ in dimensionless units in the experiment of Ref.~\cite{MFarXiv2020}, which is enough for one to observe the high values of DTOP reported in Fig.~\ref{fig:DTOP1}. It is clear that the same experimental setting can also be used to detect the Floquet DQPTs predicted in this work.

Putting together, our discoveries uncover the nonanalytic, geometric
and topological nature of Floquet DQPTs in periodically quenched systems, which not only go beyond
the conventional studies of DQPTs in sudden-quench and slow-quench
settings, but also emphasize the power of Floquet engineering in controlling
DQPTs and the usefulness of DQPTs in detecting topological phases and phase transitions
in Floquet systems. In future work, it would be interesting to extend
our framework to Floquet systems with more complicated driving protocols, non-Hermitian Floquet systems~\cite{DQPT3,NHFDQPT1,NHFDQPT2,NHFDQPT3,NHFDQPT4,NHFDQPT5,NHFDQPT6,NHFDQPT7,NHFDQPT8}, multiband systems~\cite{DQPTMultiBand}, many-particle
Floquet systems with disorder and many-body interactions, and
checking the existence and robustness of Floquet DQPTs in these complicated
settings.

\section*{Acknowledgement}
L.Z. is supported by the National Natural Science Foundation of China (Grant No.~11905211), the China Postdoctoral Science Foundation (Grant No.~2019M662444), the Fundamental Research Funds for the Central Universities (Grant No.~841912009), the Young Talents Project at Ocean University of China (Grant No.~861801013196), and the Applied Research Project of Postdoctoral Fellows in Qingdao (Grant No.~861905040009).


\end{document}